IAA-CU-13-12-01

# The Design of a Drag-Free CubeSat and the Housing for its Gravitational Reference Sensor

*Carlo Zanoni[1] [A] [B], Abdul Alfauwaz[2], Ahmad Aljadaan[2], Salman Althubiti[2], Karthik Balakrishnan[1], Sasha Buchman[1], Robert L. Byer[1], John W. Conklin[3], Grant D. Cutler[1], Dan DeBra[1], Eric Hultgren[1], John A. Lipa[1], Shailendhar Saraf[1], Andreas Zoellner[1]*

A Drag-Free CubeSat mission has been proposed to demonstrate the feasibility of a Gravitational Reference Sensor (GRS) with an optical readout for a 3 units (3U) spacecraft. A purely drag-free object is defined by the absence of all external forces other than gravity. The Drag-Free CubeSat is designed to shield a 25.4 mm spherical test mass (TM) from external non-gravitational forces and to minimize the effect of internal generated disturbances. The position of the TM relative to the spacecraft is then sensed by the GRS with an advanced LED-based Differential Optical Shadow Sensor (DOSS). This position is used in a control system to command a micro-propulsion system and to constrain the CubeSat orbit to that of the TM. In principle, the TM is then freed of all forces but gravity and the hosting spacecraft also follows a purely geodesic orbit. However, the purity of the orbit depends on the spacecraft's capacity to protect the TM from disturbances. Several of them are passively reduced by the design of the TM housing. This system is a thick-walled aluminum box that holds the shadow sensors and shields the TM. The housing has an effect on the mechanical, thermal and magnetic environment around the TM. All of them have been analyzed. The mechanical vibrations have to fit the launch environment and the modes have to be outside of the measurement range ($10^{-4}$ – 1 Hz). The magnetic field has to be reduced by a 0.01 factor. The temperature difference between internal opposing surfaces, determining pressure on the TM, has to be below $10^{-3} \times (1\ \text{mHz}/f)^{1/3}$ K Hz$^{-1/2}$. The housing, together with the TM, the sensors and the UV LEDs for charging control, constitutes the GRS, which would then fit into a 1U. The other 2Us are occupied by the caging mechanism that constraints the TM during launch, the thrusters, the Attitude Determination And Control System (ADACS) and the electronics. Next generation GRS technology for navigation, earth science, fundamental physics, and astrophysics has been under development at Stanford University since 2004. The Drag-Free CubeSat will be the result of the combined efforts of Stanford, University of Florida, KACST and NASA and will be the first drag-free mission with an optical readout and the first GRS designed within the limits of a 3U small satellite. In the first section, this paper briefly updates on the main characteristics and systems of the project. Particular emphasis is then given to the recently designed housing, its expected performance and the open issues.

**Introduction**

A free-falling body is an object whose trajectory is not influenced by atmospheric drag, solar pressure, electromagnetic interactions and any other force that is not gravity. Several space missions require a pure drag-free object inside the spacecraft in order to provide an inertial reference for the scientific phase and/or the satellite control. The heart of missions of this kind is a system called Gravitational Reference Sensor (GRS) that measures the position of a free falling test mass (TM) relative to the spacecraft. The TM position is used in a control system to command the thrusters and to constrain the spacecraft orbit to that of the TM (Figure 1). If the TM is freed of all forces but gravity, the hosting spacecraft also follows a purely geodesic orbit. The practical obstacle in obtaining a pure free-falling status is the presence of disturbances acting on the TM. These disturbances are the result of external

[1] Hansen Experimental Physics Laboratory, Stanford, CA 94305-4085, USA
[A] Dept. of Industrial Engineering, University of Trento, I-38122 TN, Italy
[B] carlo.zanoni@alumni.unitn.it
[2] King Abdulaziz City for Science and Technology, PO Box 6086, Riyadh 11442, Saudi Arabia
[3] Dept. of Mechanical and Aerospace Engineering, University of Florida, Gainesville, FL 32611, USA

interactions (air drag, solar pressure, Earth magnetic field) and spacecraft-generated effects (gravitational attraction, mechanical, electrical and thermal noise). Goal of the GRS is not only to measure the TM position, but also to shield and minimize these disturbances.

Several drag-free missions have been flown or are planned: TRIAD I in 1972, Gravity Probe B in 2004, GOCE in 2009 [1] and LISA-Pathfinder [2] in 2014 (plan). The range of applications of drag-free technology in space is wide: autonomous orbit control and maintenance, fuel and operational cost savings [3], accurate mapping of the static and time varying components of Earth's Geoid, prediction and study of catastrophic events due to scale change in water distribution, study of gravity and ultimately listen to the universe through the detection and analysis of gravitational waves. This last theoretical application "has the potential to transform much of physics and astronomy. […]and to reshape the science questions of the future" [4].

**The Drag-Free CubeSat Project**

A CubeSat with the goal of pushing these technologies has been proposed by a team composed by Stanford, University of Florida, KACST, NASA and with international support. LISA-Pathfinder is going to become the state of the art in the drag-free field. Its goal is the demonstration of the possibility of reducing the disturbances to $3\times10^{-14}$ m/s$^2$ Hz$^{-1/2}$ in the 1–30 mHz measurement band [5,6]. In the scope of the Drag-Free CubeSat such a tight requirement is relaxed to $10^{-12}$ m/s$^2$ Hz$^{-1/2}$ in the $10^{-4} - 1$ Hz measurement range. However, this mission will be the first drag-free one with an optical readout and the first GRS designed within the limits of a 3U CubeSat.

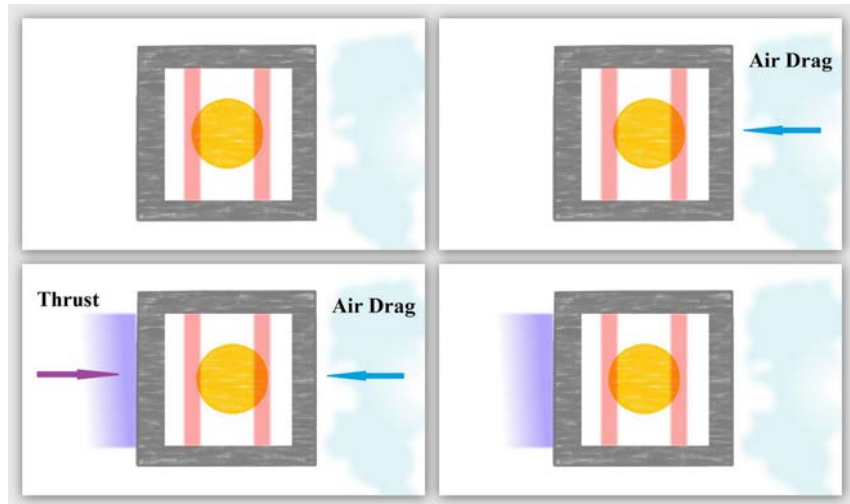

Figure 1: schematization of drag compensation

The performance requirement is a challenge for a spacecraft of such reduced dimensions (3U). The development of the Modular Gravitational Reference Sensor (MGRS), performed at Stanford since 2004 [7], provides an invaluable background for the design of this CubeSat. The primary components of the MGRS include a spherical TM, a LED based Differential Optical Shadow Sensor (DOSS), a caging mechanism based on the flight-proven DISCOS design [8] and a charge control system. At the same time, one of the key systems of the GRS in shielding from the disturbances is the housing. This is a mechanical device designed to physically hold the sensors, limit the effect of vibrations, passively reduce magnetic fields and thermal pressure and minimize the gravitational attraction induced by the spacecraft itself.

In the next section of this paper, an updated overview of the main components of the Drag-Free CubeSat is given. For a more detailed description of the mission the main reference is [9]. The remaining part of the paper is focused on the design of the housing with its performance and issues.

## CubeSat Main Components

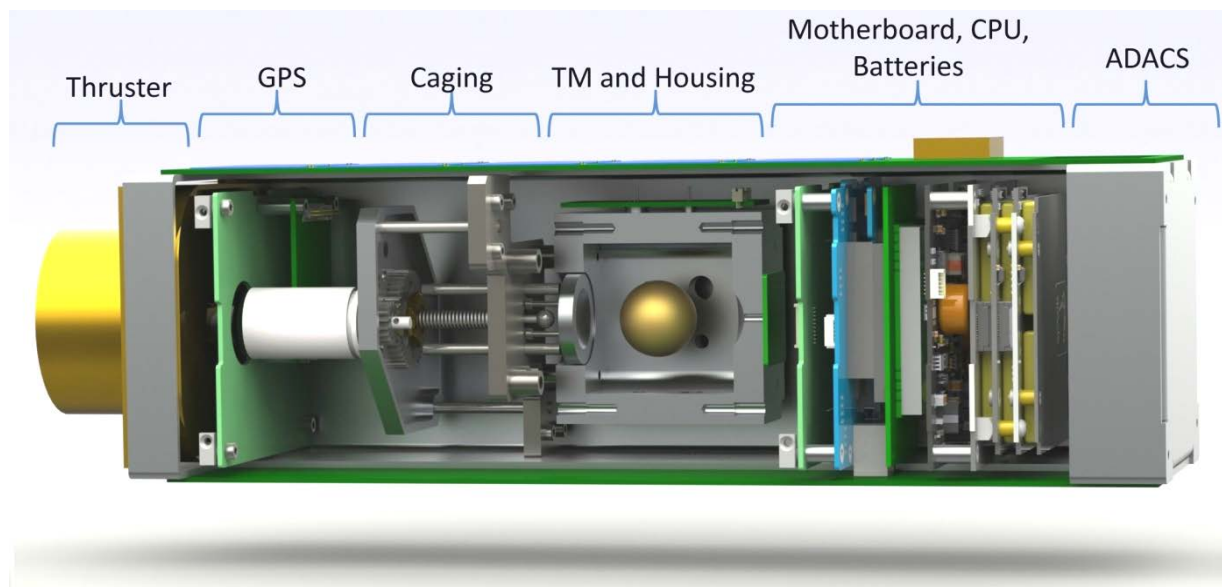

Figure 2: open view of the CubeSat

**Spacecraft Structure and Deployer**: The satellite (Figure 1) is a 3U CubeSat. All the systems are directly supported by the shell or by a Ti bulkhead normal to the satellite main axis. In the current plan, launch is assumed to be as a secondary payload in compliance with the Poly Picosatellite Orbital Deployer (P-POD) launcher (volume: 10×10×34 cm, mass: < 4 kg).

**Test Mass**: The TM is a sphere with 25.4 mm (1 inch) of diameter and 171 g of 70%/30% Au/Pt. This alloy is chosen because it is dense, can be machined, and has a low magnetic susceptibility [9]. The TM surface is coated in SiC, which has high photo-emissivity, for charge control, high elastic modulus and hardness. It is hence unlikely to obtain large adhesion forces [10], such as those measured for LISA Pathfinder [11], even after the high preloads required for launch.

**Differential Optical Shadow Sensor**: The Differential Optical Shadow Sensor (DOSS) measures the position of the TM with respect to the housing. It consists of one processing board and four identical electronic boards with two LEDs and two photo diodes each. The four boards are mounted to the housing. A total of 8 beams are centered around the TM such that half of each beam is blocked when the TM is at its nominal position. When the TM moves with respect to the housing, an intensity change is detected with the photo diodes. In order to reduce common mode noise, the difference between the measurements of two diodes on opposite sides is taken. The processing board consists of a high resolution analog-to-digital converter (ADC) and a digital signal processor (DSP). The ADC samples all 8 channels synchronously in order to reduce time jitter noise on the differential measurement. In order to improve the signal to noise ratio, especially at low frequencies, lock-in detection is used. The intensity of the LEDs is modulated at a high frequency and the DSP performs the lock-in detection.
The design goal for the DOSS is a sensitivity of 1 nm at 1 mHz.

**Caging System**: The impact constant is defined as the product of launcher acceleration, unconstrained mass and gap length. The high value of this case (~0.002 kg m) makes the presence of a caging mechanism mandatory [12]. This system (Figure 3) is designed to restrain the TM during launch and release it afterwards. It is loosely based on the design for the DISCOS system aboard the Triad satellite, which used a lead screw and plunger to restrain the spherical TM [8]. The TM is seated in a hemispherical recess on the interior wall of the housing. The system passively applies a force of at least 200 N to the TM, equivalent to the LISA requirement of 3000 N [13], scaled down by mass (1.96 kg to 0.171 kg).

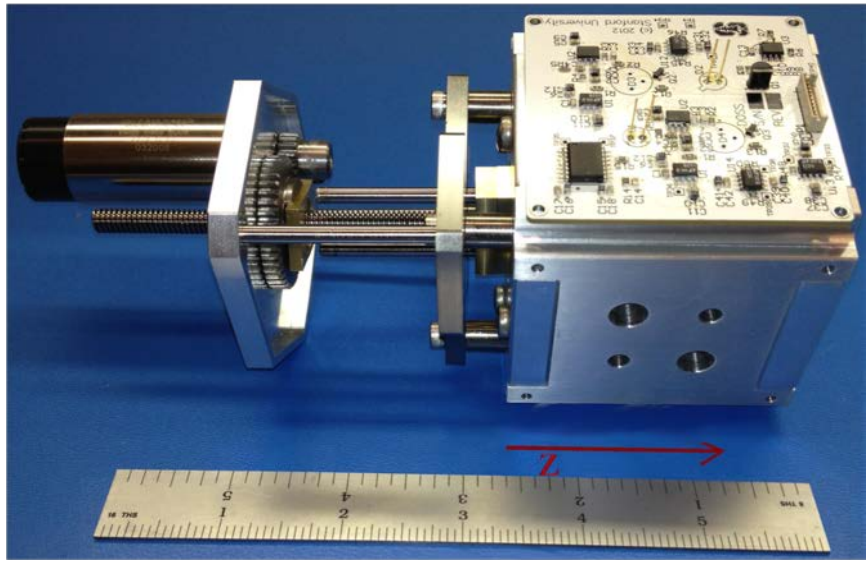

Figure 3: caging and housing (with a DOSS board). The ruler is in inches.

The caging system is fixed to the spacecraft structure by a titanium bulkhead. Titanium is chosen because it is light and strong. A fine pitch 1/4-20 acme lead screw is driven by a DC motor through a bronze nut in the bulkhead. In the next development, the drive motor will be oriented perpendicular to the direction of travel to allow more space inside the spacecraft. The total travel is approximately 25 mm, enough to move the plunger on the end of the lead screw from the locking position. Although the acme screw should prevent back-driving under most conditions, random vibration testing is planned to verify the pre-load remaining at or above the required load of 200N while the caging system is unpowered.

**UV Charge Control**: Charge imbalances between the housing and the TM caused by caging/uncaging as well the penetration of high energy particles leading to primary and secondary electron emission can cause an electrostatic disturbance force. Typical charging rates for drag-free missions are on the order of 50 electrons per second [14], with the exact rate depending on TM cross sectional area, thickness and composition of the housing, and orbit, among other factors. TM charge control is achieved via UV photoemission. Deep UV LEDs operating at 255 nm have been identified as an ideal candidate for charge control due to their small size, high dynamic range in power output, low power requirements, and the ability to be modulated at high frequency outside the drag-free control band [15]. The TM is coated with SiC because of the material's relatively high quantum efficiency of 4.86 eV. Charge control is performed via bias-free charge control: UV light is shined onto the TM, and some light reflects back to the housing. Photoelectrons are generated from both surfaces and the electrons will preferentially move such that the potential between the two surfaces reaches equilibrium.

**Drag-Free and Attitude Control System**: The Drag-Free and Attitude Control System (DFACS) is the system that constraints the orbit of the spacecraft to that of the TM. 6 degrees of freedom (DOF) are sensed and 4 are actuated. The thruster applies its force along z. The actuation from a cold-gas thruster has to be an on/off one. With the on/off actuation, the motion of the satellite is a sequence of parabolic arcs (~$4\times10^{-4}$ m in amplitude). The limit of this algorithm is the minimum impulse bit of the thruster. A reduction of this parameter would allow better drag-free performance.

Atmospheric drag is not necessarily the only external disturbance force. The total force won't be aligned with the direction of travel and attitude control is required in the compensation of these non-gravitational actions. Attitude actuation is a task of the Attitude Determination and Control System (ADACS) at the top (+z) of the satellite. The main x y and z sensors are still the DOSS and a set of IMU accelerometers provide back-up information. Attitude sensing is a fusion of the ADACS horizon sensor and the IMU rate gyros [9]. Drag-free control in the transverse x and y directions is performed by adjusting the attitude such that the CubeSat opposes the drag force. The expected yaw and pitch angles are < 10 deg.

Finally, as no differential measurement between test masses is used and the TM is a sphere, electrostatic actuation inside the housing is not required. This allows substantial savings in complexity, money and mass.

**Micro-propulsion**: The preliminary requirements for the micro-thrusters are an impulse of between 10 mNs and 100 mNs with a precision of better than 1 mN·s. As baseline a micro propulsion cold gas thruster by VACCO is used. This thruster has axial thrust capability only. A number of other options are being considered, including newer versions of the cold gas thrusters by VACCO and other companies, ionic fluid (also known as colloidal) thrusters and field emission thrusters.

**The Housing for the Gravitational Reference Sensor**

The mechanical part of the GRS is constituted by the housing, which is a box surrounding the TM. Its main functions are holding the sensors and other GRS systems and passively reduce the effects with a negative impact on the drag-free performance. Compared to LISA-Pathfinder, the GRS complexity is here reduced by the absence of electrostatic sensing and control inside the housing.

   *1. Requirements*

In order to comply with the drag-free performance goal, the housing has to:
   1. hold, together with the caging mechanism, the TM during launch;
   2. allow a safe release;
   3. shield the TM thermally: maximum temperature difference between internal opposing surfaces below $10^{-3} \times (10^{-3}/f)^{1/3}$ K Hz$^{-1/2}$;
   4. shield the TM magnetically, with a 0.01 reduction factor;
   5. hold the shadow sensors rigidly: the mechanical modes of the housing inside the spacecraft have to be clearly outside of the measurement range ($10^{-4}$ – 1 Hz);
   6. minimize the gravitational gradient;
   7. make a re-caging possible;
   8. hold the UV LEDs for charge control;
   9. have a surface that allows charge control;
   10. minimize patch effects;
   11. fit both the DF CubeSat and the ShadowSat (the CubeSat for DOSS testing, [16]) with minimal re-design;
   12. be machinable;
   13. be vacuum compatible;
   14. be easily assemblable for tests;
   15. be compatible with the systems already designed (Caging, DOSS, CubeSatKit...).

Some of these requirements are guaranteed by deposing a coating inside the housing. More specifically:
   1. a safe release is allowed by minimizing the adhesion bonding between TM and housing. This means that the TM does not remain attached to the caging or the housing when the first one is retracted by a small amount. At the same time, the TM residual velocity is high after the release (it's an effect of the 200 N preload). This velocity is then dissipated with a certain amount of bounces of the TM between the caging and the housing. Therefore, the geometry has also to survive these bounces and especially avoid small damages to the TM. The surfaces have then to be hard and chemically inert;
   2. charge control depends on the photo-emissivity of the TM and of the housing surfaces where the UV rays are reflected;
   3. patch effect is minimized by a short wavelength voltage variability on the inner surfaces.

A material whose properties fit these guidelines is the SiC (Silicon Carbide) that covers the TM too. However, the use of a coating inside the housing imposes tight constraints on the geometry of the housing itself because the coating deposition process has to be taken into account. A uniform distribution is allowed only with an open geometry. The best would be the separation of all the sides in different pieces. On the other hand this would make the tolerancing and alignment much more critical.
It is then worth noting that the spherical TM requires no forcing or orientation control, and allows for larger TM-to-housing gaps. At the same time, also patch effect strongly depends on, and thus limited by, this gap. For this reasons, the housing dimensions have to be larger than the TM radius allowing a big TM-to-housing gap.
At this point, it is worth noting that the minimization of adhesion force and charge control require the presence of coating only on the surface around the UV LEDs and where the TM is held during the launch. As a consequence, assuming that the housing is a cube, the surfaces primarily interested by the coating are only the two normal to the CubeSat main axis (z, in Figure 2).

## 2. *Final Geometry*

It is highly preferable that the piece supporting the DOSS is not fractioned allowing better sensors alignment. A choice has been made between uniform coating deposition on every surface and improved tolerancing for sensor alignment. As a consequence, The final housing design has to allow coating deposition on the +-z inner surfaces. Besides, large gaps to the TM are required for improved minimization of disturbances.

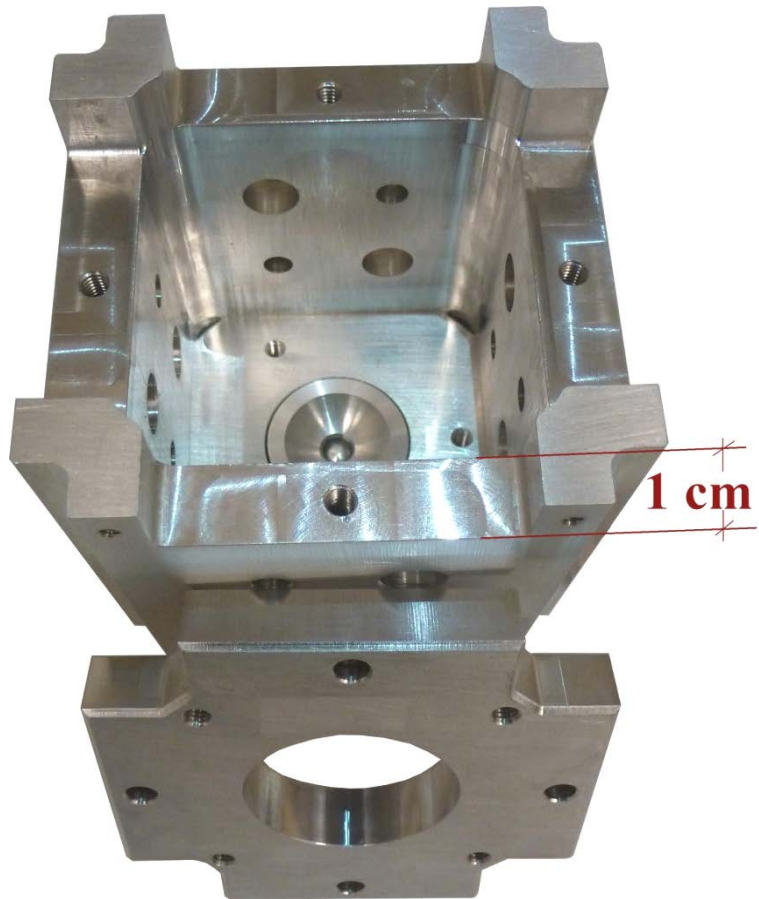

Figure 4: housing model for shaking test. The hemispherical recess on the bottom is here connected to a load cell.

Taking all this into consideration, the final geometry is a thick-walled cube with 70 mm external and 50 mm internal side. This cube is divided in 3 parts (Figure 4 and 5), + z and –z caps and the remaining sides (as a tube). These two caps will be covered with SiC on the inside. To prevent extreme events (e.g. TM bonded to a side surface), it is still possible to depose coating on the other surfaces, but with less control on the thickness and reduced uniformity. The + z cap has to include a hemispherical recess that holds the TM during launch. The –z part has to allow the motion of the caging mechanism. The material chosen is Al6061, or Al7075, that has relatively low density, high yielding stress and good thermal conductivity. The x and y sides of the housing have holes for DOSS LEDs and diodes. The outer edges along z are cut to host the linear guides of the caging mechanism. The 3 parts are then assembled together with a set of screws (M4). The resulting box is connected to the caging and is mounted on a titanium bulkhead perpendicular to the CubeSat z axis (Figure 2).

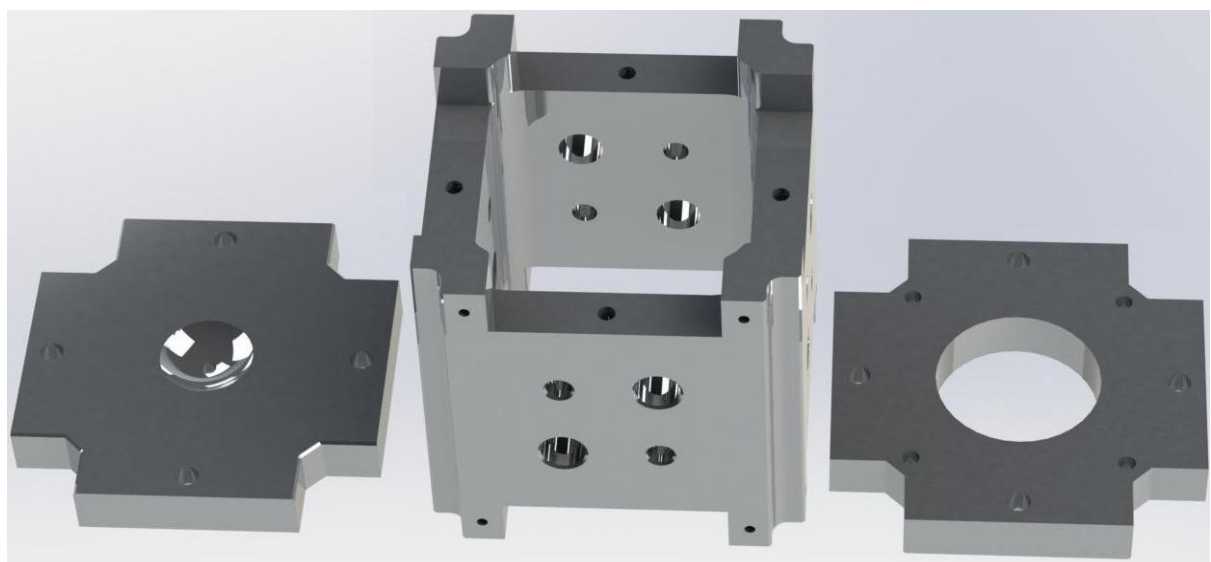

Figure 5: CAD model of the disassembled housing

Figure **1** shows the modular structure of this design of the housing. The +- z caps can be substituted with different pieces when required for testing purposes (e.g. a load cell mounted to the hemispherical recess during the shaking test).
An important source of disturbance is the gravitational force induced by the presence of the spacecraft itself. The minimization of this source could impose slight changes in the final geometry. However, this effect will be analyzed when the project is in a more advanced state because the final position and the mass properties of all the devices have to be precisely known. At that point, the spacecraft center of mass has to be located on the TM center. Also the gravitational gradient has an impact on the disturbance (the TM is not continuously in the center). In order to minimize this value, the inertia of the spacecraft has to be considered [17]. However, as a general guideline, a good GRS design is always the most symmetrical. That is why, for instance, all the corners of the housing are cut even if only two of them host the caging linear guides.

### 3. *Mechanical Analysis*

The housing has to survive the launch environment, allow a safe caging and release of the TM and hold the sensor rigidly enough. While the best test to identify and qualify a complex mechanical system with several parts and screws is a shaking test, an estimation of the expected behavior has been performed with a FE analysis in COMSOL.
The modes of both the housing alone and the housing mounted in the spacecraft have been computed. In order to bound the numerical error a convergence analysis has been performed. The results (Table 1) do not suggest any kind of problem. However, it will be useful to perform a comparison between numerical and measured values. As expected, the behavior of the housing alone is symmetrical.

| Mode | Housing alone (Hz) | Housing mounted inside the spacecraft (Hz) |
|---|---|---|
| 1st | 2596 | 227 |
| 2nd | 2598 | 391 |
| 3rd | 5920 | 497 |
| 4th | 6112 | 628 |
| 5th | 9519 | 695 |
| 6th | 9525 | 719 |

Table 1. modal analysis results

The static analysis of the housing with a caged TM (200 N of load) provides a maximum Von Mises stress of 27.5 MPa (safety factor: 11 and 16.5 for Al6061 and Al7075 respectively). This value does not take into account dynamic effects. On the other hand, this analysis provides precious information on the most critical features of the housing (Figure 6). The highest stress is located where the housing is fastened with the bulkhead (and the caging).

The planned shaking test will provide the final word on the design, on its capability to survive the launch environment and on potential damages on the TM and housing surface. The housing for the shaking test is slightly different from the proposed flight geometry. As a matter of facts, a load cell will be mounted on the +z cap to check the loads before, during and after the test.

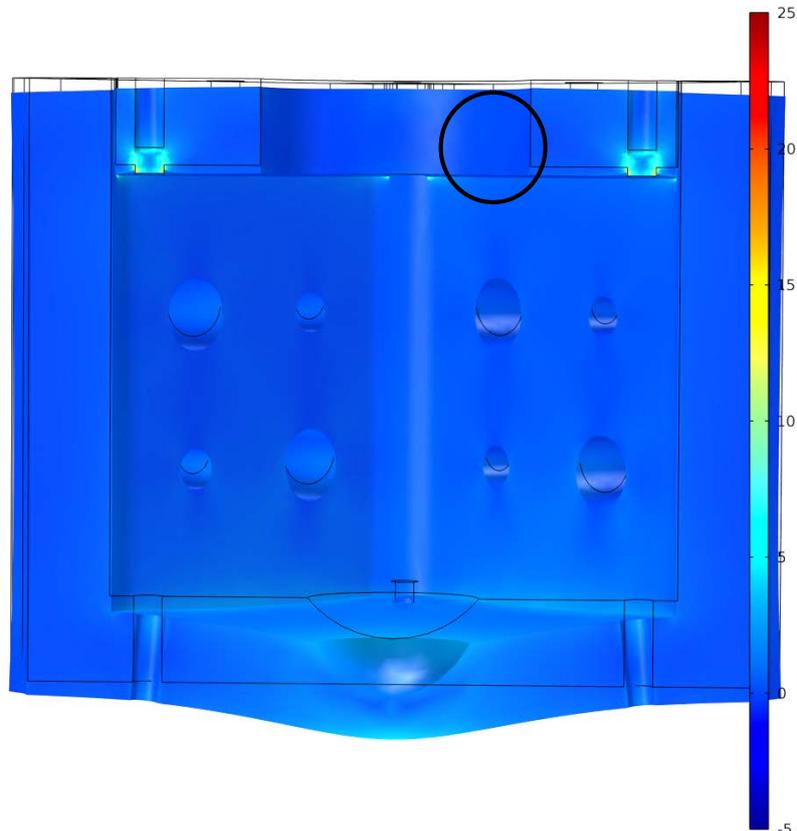

Figure 6: Von Mises stress and the deformation (enhanced) of the housing in caged static status.

### *4. Magnetic Analysis*

The preferable spacecraft orbit is a sun-synchronous LEO. This means the satellite (as it happens for every CubeSat) will be close to Earth and its environment. One major effect is then the magnetic field that could induce on the TM. Also some systems inside the spacecraft generate a magnetic field. The housing has a major role in reducing this effect (reduction factor between the field inside and outside the housing: 0.01). In order to reach this goal a thin cover of a specific material has to be applied on the housing aluminum box. A FE analysis has been performed in Ansoft Maxwell to compare two materials (mu-metal and 2714A) and find the thickness required.

The geometry used in the analysis is a simplified one (no small features, no caging). Besides, the shielding is modeled as a uniform cover around the housing. So far, no real design on how this material is held has been defined. The excitation is a magnetic field with the highest value of the Earth magnetic field. Several analyses changing the angle between the field and the spacecraft have been performed. The reduction factor is defined as the ratio between the average field inside the housing and the boundary excitation.

The results show a 10X better behavior of 2741A compared to mu-metal (Figure 7). The reason for this behavior is the extremely high permeability of 2741A. In any case, the shielding material will be thinner than 0.1 mm. This means a foil, instead of a rigid plate, is enough to comply with the requirement.

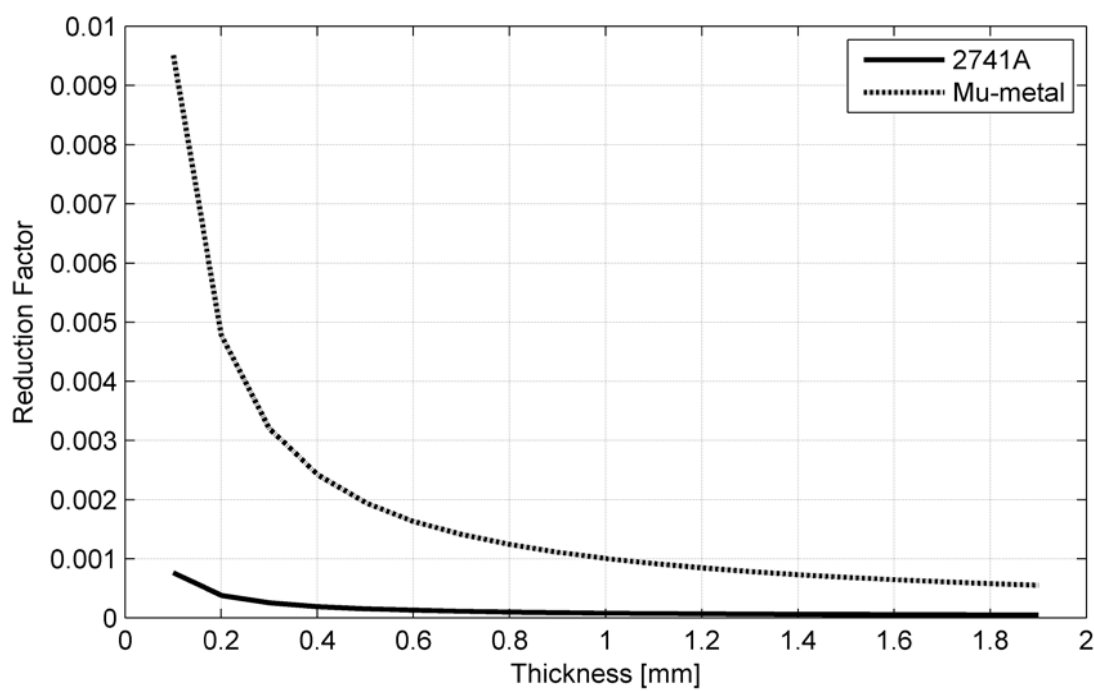

Figure 7: reduction factor as a function of the thickness of the shielding material (worst case angle between spacecraft and field, i.e. field parallel to z)

### *5. Thermal Analysis*

Thermal environment is one of the most critical aspects for every satellite. The drag-free performance requirement enhances this aspect. A temperature difference between opposed internal faces of the housing acts on the residual gas around the TM and results in a net force on the TM. In the current requirement, the maximum temperature difference is set to $10^{-3} \times (1 \text{ mHz}/f)^{1/3}$ K Hz$^{-1/2}$.

FE analyses have been set up in COMSOL for this thermal problem. Accuracy is critical. The geometry of the model is therefore comprehensive of most of the mechanical features and power generating electronics.

These are the main aspects of the simulation (Figure 8):
1. FE model includes spacecraft shell, thrusters, bulkhead, caging, housing, TM and DOSS

   electronics;
2. entering heat flux: Sun (1000 W/m$^2$), Earth (400 W/m$^2$);
3. outgoing heat flux: radiation with $T_{amb}$ = 4 K;
4. internal heat generation: 1 W (electronics);
5. emissivity external surfaces: 0.7 (averaged between solar panels and spacecraft aluminum);
6. internal radiation and conduction allowed.

The values and assumptions for the simulations are derived from [18].

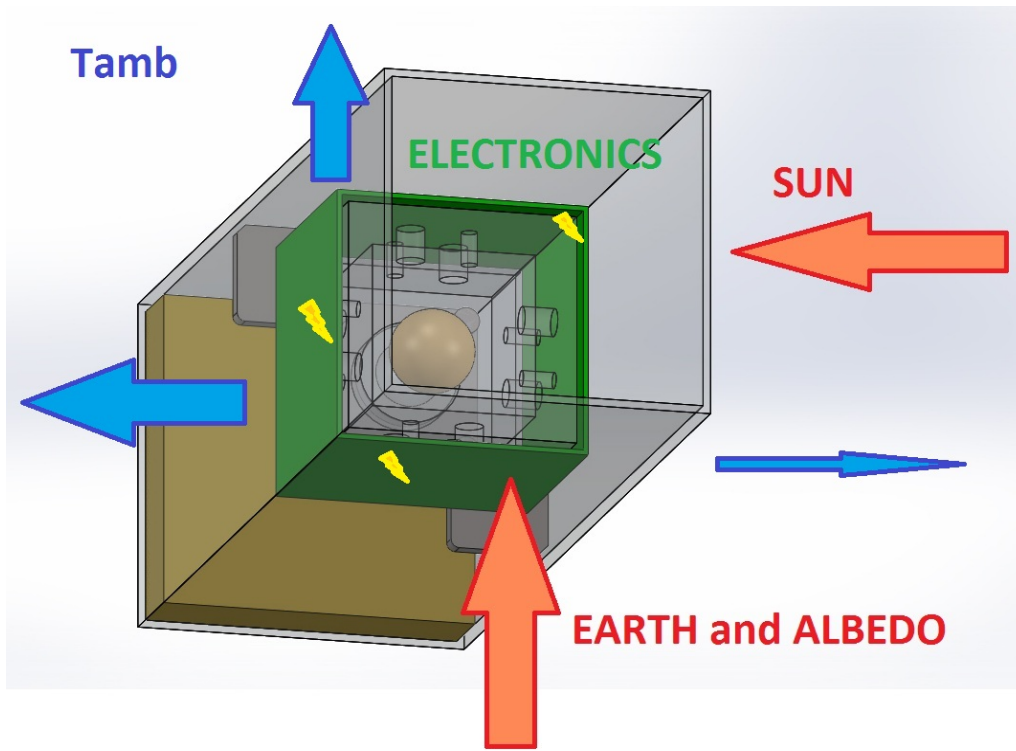

Figure 8: schematic of the FE thermal model

In the first analysis the model is subjected to step excitations (heat flux, power generation…) from the nominal conditions (293 K). The result provides an estimation of the performance in case of eclipse. This analysis has also the purpose of validating the model thanks to a comparison with a 1 degree of freedom analytical model. The steady-state temperature on the external sunny (hot) side is different by less than 3 K despite the assumptions for the analytical model (no 3D geometry, the spacecraft is a unique object).

The result of the simulation has then been analyzed in the frequency domain and compared with the requirement (Figure 9). The temperature difference inside the housing along z is clearly above the requirement limits.

However, a more realistic analysis is that one simulating the effect of a heat flux with a harmonic term. More specifically, the Earth heat flux contributes with a time-dependent component $A\sin(2\pi f\, t)$, where $A$ has been used both with 80 W/m$^2$ and 40 W/m$^2$ value. $f$ is the frequency and changes value in each simulation. However, in order to keep the FE analysis on a simple level, the maximum frequency has been $2\times10^{-3}$ Hz. Otherwise, the period of the harmonic term and the time-constant of the spacecraft would be too different, requiring lengthy simulations. A certain amount of simulated time is required before the system reaches a steady state harmonic response if the frequency is high the number of needed steps becomes higher.

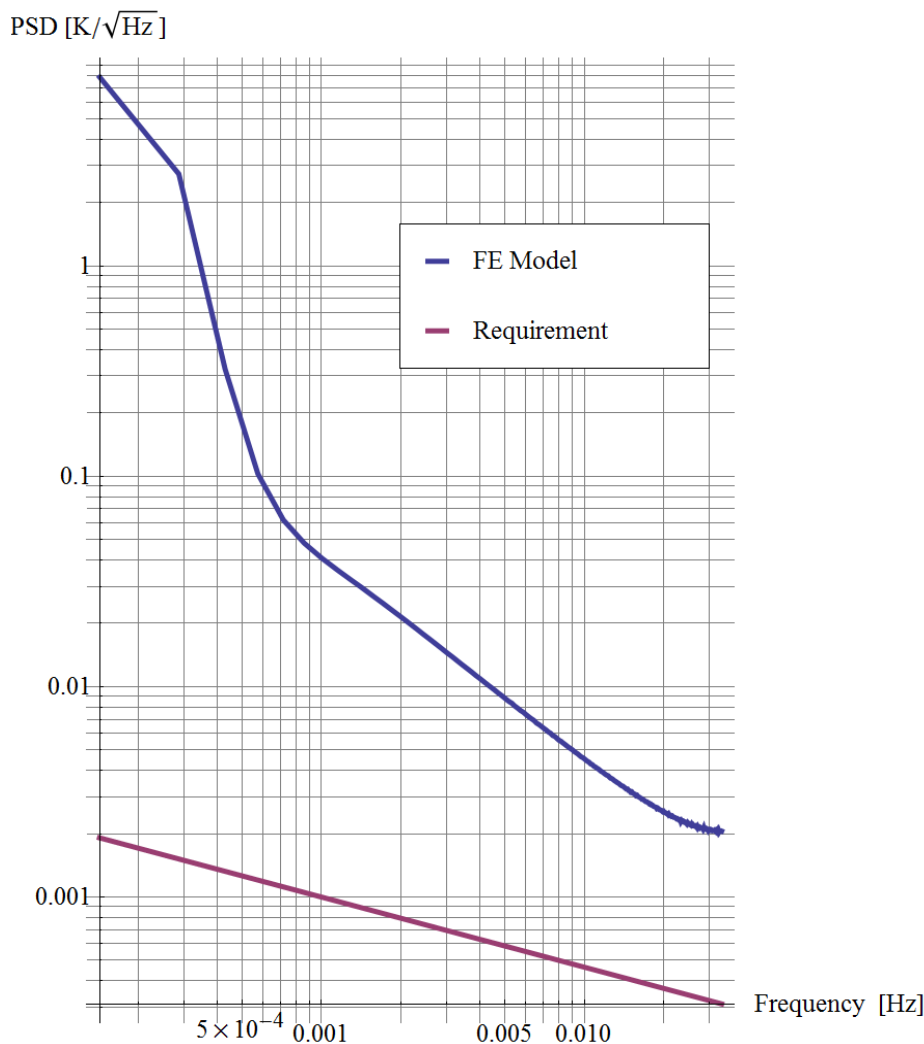

Figure 9: PSD of the temperature difference transient along z inside the housing (legend from top to bottom)

The thermal performance is clearly above the requirement limits. Several reasons have been identified for this behavior:
1. the geometry of the housing is not symmetrical as on one side there is the caging which results in a different thermal path and inertia;
2. the requirement alone is particularly demanding. For instance at $10^{-4}$ Hz the maximum temperature difference is $2.15 \times 10^{-5}$ K;
3. the FE simulation requires a large number of elements and nodes on a complex geometry. This hinders the accuracy of the model.

This problem has to be understood and solved. The first step will be an improvement in the analysis with software more specific for thermal and space problems. Particular care has to be taken in the definition of the mesh. The requirement has then to be analyzed and possibly relaxed. Finally, with a better understanding, a more symmetrical and better geometry has to be suggested. A preliminary analysis with a housing supported on two symmetrical bulkheads has been run. The performance improves but just by a factor 2 or 3. The reason for this is the presence of the Caging System. Its presence alone is a strong violation of symmetry. A more complex solution has then to be found. Once a promising geometry and configuration is found, experimental measurements should be performed to validate the thermal design.

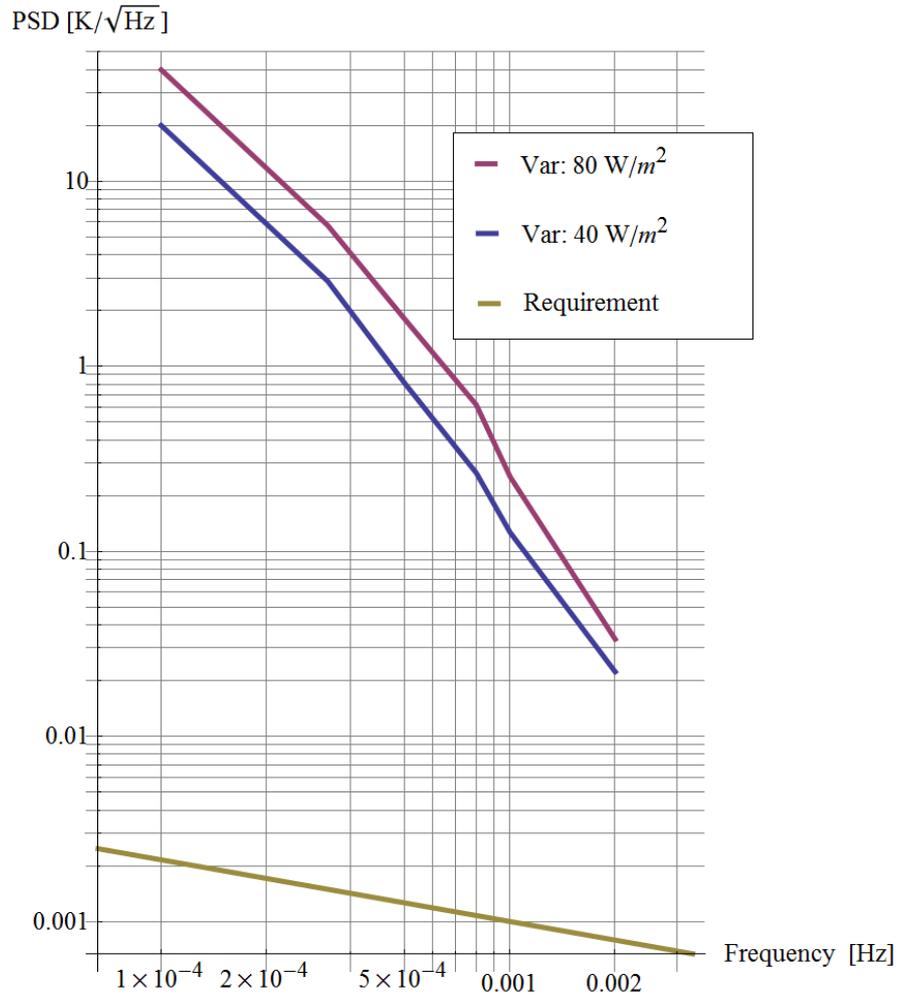

Figure 10: PSD of the behavior of the temperature difference inside the housing (legend from top to bottom)

## Conclusions

Drag-free technology has recently seen increased interest in both European and US space agencies. The range of applications is wide and goes from orbit control and maintenance to the study of the fundamental laws governing the universe. The proposed Drag-Free CubeSat is described here in its main features. Key factor in the success of a space mission of this kind is the limitation of disturbances acting on the test mass that is free-floating inside the spacecraft. One of the systems involved in the disturbance reduction is the housing that hosts the TM, supports the sensors and shields from several disturbances. The current design and the mechanical, magnetic and thermal analyses performed on this system are described here. The results show that the thermal performance does not appear to be compliant with the requirement. This is an indication of the Drag-Free CubeSat being a huge technological challenge. As a matter of facts, this satellite will be the first drag-free mission with an optical readout and the first GRS designed within the limits of a 3U CubeSat.

## Acknowledgements

This project is supported by the KACST Center of Research Excellence in Aeronautics and Astronautics (Saudi Arabia) at Stanford University. Carlo Zanoni is grateful to Prof. J.W. Conklin, the SpaceGrav group at the Hansen Experimental Physics Laboratory and the University of Trento for supporting his stay in Stanford. He would also like to thank Prof. D. Bortoluzzi (University of Trento) for his useful comments on this paper. The presentation of this work at the 2nd IAA Conference on University Satellites Missions and CubeSat Workshop is kindly sponsored by the European Space Agency (ESA).